\documentstyle[12pt]{article}
\begin{document}
\textheight 200mm
\textwidth 160mm
\leftmargin 10mm
\newcommand{\be}{\begin{eqnarray}}
\newcommand{\ee}{\end{eqnarray}}
\newcommand{\zt}{\zeta}
\newcommand{\ve}{\epsilon}
\newcommand{\al}{\alpha}
\newcommand{\gm}{\gamma}
\newcommand{\bt}{\beta}
\newcommand{\dt}{\delta}
\newcommand{\la}{\lambda}
\newcommand{\vp}{\varphi}
\newcommand{\nn}{\nonumber}
\renewcommand{\baselinestretch}{1.4}
\begin{center}
 { \Large\bf
 New approach to Borel summation of divergent series and critical
 exponent estimates for an $N$-vector cubic model in three
 dimensions from five-loop  $\ve$ expansions}
\end{center}
\vspace{0.3cm}
\begin{center}
 {\bf A. I. Mudrov }
\end{center}
\begin{center}
Department of Theoretical Physics,
Institute of Physics, St.Petersburg State University, Ulyanovskaya 1,
Stary Petergof, St.Petersburg, 198904, Russia\\
e-mail: aimudrov@dg2062.spb.edu
\end{center}
\begin{center}
{\bf K. B. Varnashev}
\end{center}
\begin{center}
Department of Physical Electronics, Electrotechnical University, \\
Prof. Popov Street  5, St.Petersburg, 197376, Russia
\end{center}

{\footnotesize
A new approach to summation of divergent field-theoretical series is
suggested. It is based on the Borel transformation combined with a
conformal mapping and does not imply the exact asymptotic parameters
to be known. The method is tested on functions expanded in their
asymptotic power series. It is applied to estimating the critical
exponent values for an $N$-vector field model, describing
magnetic and structural phase transitions in cubic and tetragonal
crystals, from five-loop $\ve$ expansions.}

\vspace{0.5cm}
\vspace{0.5cm}

Recently Kleinert and Schulte-Frohlinde calculated renormalization
group (RG) functions for the  cubic model in ($4-\ve$)-dimensions
within the five-loop approximation of the renormalized
perturbation theory \cite{1}. The critical (marginal) dimensionality
$N_c$ of the order parameter field found on the base of those
expansions turned out to be smaller then three in three dimensions.
This means that the critical behavior of the model should be
controlled by the cubic
fixed point with a specific set of critical exponents. Thus,
calculation of the critical exponent values for the cubic model is an
actual problem. Since the model of concern describes
the critical behavior of cubic and tetragonal crystals
undergoing magnetic and structural phase transitions, critical
exponent estimates can be compared with experimental data.

The field-theoretical RG series are known to be badly divergent.
To extract the physical information relevant to predicting the critical
behavior of real systems they should be processed by a proper
resummation procedure. The series for $N_c$ obtained in Ref. \cite{1}
proved to be alternating in signs that allowed to resumm it
by the simple Pad\'e method. At the same time, the critical exponent series
are irregular and cannot be processed by the ordinary
(Pad\'e, Pad\'e-Borel, Pad\'e-Borel-Leroy) techniques.
A more sofisticated method of the Borel transformation combined
with a conformal mapping, although regarded as the most universal
procedure, is inapplicable as well, because it requires to know
the exact values of the asymptotic parameters characterizing the
high order behavior of the series. Nowadays those parameters have been
evaluated for the simplest case of the $O(N)$-symmetric models
only \cite{2,3}, and calculating them for anisotropic models is
a most difficult problem as yet unsolved. As an exception one
should mention the anisotropic quartic quantum oscillator which represents
a one-dimensional $\varphi^4$-field theory with the cubic
anisotropy. For the perturbation expansion of the ground state energy
of this system the asymptotic parameters were found in Ref. \cite{1a}.
Within the assumption of the  weak anisotropy the large-order asymptotic
behavior of the $\beta$-functions for the cubic model  was
deduced in Ref. \cite{1b} and then used for determination of the
stability of the cubic fixed point in three dimensions \cite{1c}.

The aims of the present work are i) to suggest a new approach to
summation of the divergent field-theoretical series, which is based
on the Borel transformation combined with a conformal mapping, but
which does not involve the exact values of the asymptotic parameters.
This approach, being the result of a computer study of the
Borel summation procedure, is tested on functions
expanded in their asymptotic power series, on caclulation of
the ground state of the unharmonic oscillator, and on
estimation of the critical exponent values for the basic
models of phase transitions,
ii) to obtain critical exponent estimates for the cubic
(simplest anisotropic) model in three dimensions from
the record five-loop $\ve$ expansions \cite{1}, using the
developed technique.

We should like to emphasize that the problem of processing
divergent series arises in various fields of physics, where
the perturbation theory is employed but the parameter of expansion
is not small. So, developing a resummation procedure which
could be effective where conventional methods fail is of
general interest.

A modification of the Borel procedure via a conformal mapping was
introduced in Ref.\cite{4} and used for processing series
$\sum_k f_k g^k $, whose coefficients at large order $k$ behaved
as $k! k^{b_0} (- a_0)^k$.
To the series $\sum_k f_k g^k $ the function
\be
F(g;a,b)=
\int_0^\infty   e^{-\frac{x}{a g}} \Bigl(\frac{x}{a g}\Bigr)^b
d\Bigl(\frac{x}{a g}\Bigr) B(x)
\label{eq1}
\ee
is associated. The Borel transform $B(x)$ is the analytical continuation
of its Taylor series $\sum_k \frac{f_k}{a^k\Gamma(b+k+1)} x^k$
absolutely convergent in the unity circle. Usually $a=a_0$, whereas
the parameter $b$ may be  related to the exact asymptotic value
$b_0$ in a variety of ways \cite{4,5}. Conformal mapping
$\omega=\frac{\sqrt{x+1}-1}{\sqrt{x+1}+1}$ transforms the cut--plain
${\bf C}\backslash [-1,-\infty)$ onto the unity circle, and the
semi--axis $[0,\infty)$, the domain of integration, goes over into
the interval $[0,1)$. Supposing $B(x)$ may be continued over the
cut--plain, the composite function $B(x(\omega))$ is holomorphic
within the unity circle and its Taylor series admits integration
before summation. In order to eliminate possible singularity of
$B(x(\omega))$ at $\omega=-1$ an additional parameter  $\la$ was
introduced in Ref. \cite{5} and used in Refs. \cite{6,7}:
$B(x(\omega))=\frac{A(\la,\omega)}{(1-\omega)^{2\la}}$. It is chosen
from the condition of the most rapid convergence of the series
\be
 F(g;a,b) =\sum_{k=0}^\infty A_k(\la) \int_0^\infty
 e^{-\frac{x}{a g}} \Bigl(\frac{x}{a g}\Bigr)^b
 d\Bigl(\frac{x}{a g}\Bigr)
 \frac{\omega^k(x)}{(1-\omega^k(x))^{2\la}},
 \label{eq2}
\ee
 that is from minimizing the quantity
 $|1-\frac{F_L(g;a,b)}{F_{L-1}(g;a,b)}|$, where $L$ is the step
 of truncation and $F_L(g;a,b)$ is the $L$--partial sum  of the
 series (\ref{eq2}).
 In practice, one deals with a piece of the series only, where the
 asymptotic regime might not be reached. For this reason, in
 Refs. \cite{4,8,9} the parameter $b$ was varied in a neighborhood
 of $b_0$. We believe that in the case of the unknown exact
 asymptotic value  $a_0$ similar manipulations may apply to the
 parameter $a$ as well.

 Our approach to using the procedure described above consists in the
 following. While processing the asymptotic series of {\em a priory}
 given functions we have revealed that the quantity $F_L(g;a,b)$
 remains stable as the parameters  $a$ and $b$ vary in a wide range.
 With the "number of loops" increasing, the ($a,b$)--dependence
 becomes weaker and weaker, and the difference between $F_L(g;a,b)$
 and the exact value  $F(g)$ can be made whatever small. This
 observation enables us to employ the Borel transformation with a
 conformal mapping for processing series whose exact asymptotic
 behavior is unknown. Thus, we postulate stability of the result of
 the processing with respect to variation of $a$ and $b$ as a basic
 principle underlying our approach.

 Let us demonstrate how this principle works on some simple examples.
 At first, consider a model function
 $${\cal F}(g) = \int_{-\infty}^{+\infty} e^{-x^2-g x^4}dx
 \sim\sum_{k=0}^\infty (-1)^k \frac{\Gamma(2k+\frac{1}{2})}{k!}g^k$$
  whose coefficients $f_k$ behave as
 $\frac{(-4)^k}{\sqrt{2\pi}}\frac{k!}{k}$
 at large $k$. For each $a$ and $b$ in (\ref{eq2}) we find the set
 of values
 $\{\la^i_{min}(a,b)\}$, at which the quantity
 $|1-\frac{{\cal F}_L(g;a,b)}{{\cal F}_{L-1}(g;a,b)}|$, as a function
 of $\la$, reaches its local minimums. To each element of that set
 corresponds the particular value ${\cal F}^i_L(g;a,b)$. It can be
 shown that taking into account relative contributions of lower order
 terms of the series ($k<L$) when optimizing the $\la$--parameter
 just weakly affects the final result, therefore it is sufficient to
 minimize the relative contribution of the last accounted term only.
 At Fig.1 (a) a few curves  ${\cal F}^i_{13}(0.5;a_0,b)$ are
 presented, which, as calculations show, deviate from the exact
 number ${\cal E}(g)=1.48631082...$ by no more than $5\cdot 10^{-5}$
 within the range $0\leq b \leq 18$.

 For the function ${\cal F}(g)$ the exact asymptotic value of $a$ is
 $a_0=4$. As the parameter $a$ shifts from $a_0$ the picture does not
 change qualitatively, although at large numbers one can see a
 reduction of the $b$-stability interval, as displayed at Fig.1 (a).
 For small $a$ the set $\{\la^i_{min}(a,b)\}$ is just empty within the
 interval $[0,6]$. In all cases the lines in their stability domains
 are at the same level with the mentioned above accuracy. So, on the
 graph of ${\cal F}^i_{13}(g;a,b)$ depending on two parameters $a$ and
 $b$ there would be a horizontal part -- plateau rapidly destroyed at
 its boundaries.

 The size of the plateau depends essentially on $g$. When $g$ gets
 smaller the stability domain is enlarged. That is in accord with
 intuitive expectations that the decrease of the expansion variable
 should result in better convergence. So, the calculations show that
 for the family of five  curves ${\cal F}^i_{13}(0.1;a_0,b)$ the
 relative error
 $|\frac{{\cal F}^i_{13}(0.1;a_0,b)-{\cal F}(0.1)}{{\cal F}(0.1)}|$
 does not exceed $2\cdot 10^{-6}$ within the interval
 $0\leq b \leq 25$ (see the table in Appendix). 
 On the contrary, the growth of $g$ leads to
 reduction of the stability domain. Let us follow how its size
 changes from loop to loop, assuming $g=1$. Presented at Fig.1 (b)
 are the most stable curves ${\cal F}^i_L(1.0;a_0,b)$ for $L=6,9,12$.
 As seen from the graphs, with accounting more terms of the
 expansion the behavior of the curves improves, and the dispersion
 of  the values goes down. The graph illustrating convergence of
 numerical estimates of ${\cal F}(g)$ for $g=1$ depending on
 the truncation number $L$ is depicted at Fig.2

 When processing functions
 $\int_0^\infty e^{-x}(x\partial_x)^{b_0}\frac{1}{1+g x}dx \sim
 \sum_{k=0}^\infty (-1)^k k! k^{b_0} g^k $, where $b_0\geq 0$, we
 observed the similar stability with respect to $a$ and $b$.
 We also processed the six--loop pieces of RG series for the critical
 exponents of the $O(N)$--symmetric model in three dimensions and
 found very weak dependence of the output on the transformation
 parameters. The numerical values of the critical exponents
 computed proved to be in a good agreement with those of
 Ref. \cite{4}.

 Using the proposed summation method for processing the series of the
 ground state energy ${\cal E}(g)$ of the unharmonic
 oscillator \cite{9a}  with the Hamiltonian $H=x^2+g x^4$ we observed
 the same behavior of the corresponding curves as for the model
 functions considered above. In Table 1 we present the
 estimates of ${\cal E}(g)$ at $g=1$ depending on the
 length of the series.

\vspace{0.3cm}
 TABLE I. Numerical estimates for the unharmonic oscillator ground
 state energy at $g=1$.

\vspace{0.3cm}
\hspace{0.2cm}
\begin{tabular}{|c|c|c|c|c|c|c|}\hline
L
&         8&        9&      10&       11&       12&Exact  value
                   \\[0pt]\hline
${\cal E}$(1)
&1.392376 &1.392357 &1.392344 &1.392349 &1.392351 &1.392352
                   \\[0pt]\hline
\end{tabular}

\vspace{0.3cm}
\noindent
Note that for $L=8$ our estimate is by one order closer to the
exact value than the number $1.391655\pm 0.004562$ found in Ref.
\cite{10} on the base of Wynn's $\ve$--algorithm.

 The numerical analysis fulfilled allows to apply the summation method
 introduced to finding numerical estimates of the critical
 exponents for the $(4 - \ve)$--dimensional cubic model \cite{1}
 in three dimensions $(\ve = 1)$. The $b$--dependence of the
 results of processing the exponent $\gamma^{-1}$ for $N=2$ at
 various fixed $a$ is presented at Fig.3. The parameter
 $a$  ranges from $0.2$ to  $1.5$ with step $0.1$. Distinct
 oscillations correspond to the small values of $a$ and $b$.
 As $a$ grows, the behavior of the curves becomes smoother and
 the extended horizontal interval appears reaching its maximum
 length at about $a=0.5$. Further increase of $a$ makes the stability
 interval  getting shorter  and causes the rapid growth of the curves
 at its  boundaries. It is essential that for all values of
 $a$  the  horizontal parts of the curves are at the same level.
 Averaging the results of the processing over the stability domain
 gives the number which we adopt as $\gamma^{-1}$ sought for. The
 accuracy for this  approximate value is determined through the
 dispersion due to the  variation of $a$ and $b$. Similar behavior
 of the curves was observed for other critical exponents. Applying
 to them the developed algorithm we obtain the critical exponent
 values for the cubic model for various $N$ listed in Table 2.

\vspace{0.3cm}
 TABLE II. Critical exponents for the cubic model in three
 dimensions from the five--loop approximation in $\ve$ for
 some $N$.

\vspace{0.3cm}
\hspace{-0.5cm}
\begin{tabular}{|c|c|c|c|c|r|}\hline
N &$\eta           $ & $\nu            $&$\gm
$&$\gm_{sc}$&$\frac{\gm-\gm_{sc}}{\gm_{sc}}\cdot 100\%$
                   \\[0pt]\hline
2 &$0.0350\pm0.0003$ & $0.6277\pm0.0010$&$1.2358\pm0.0040
$&$1.2334$&0.19 \%
                   \\[0pt]\hline
3 &$0.0375\pm0.0005$ & $0.6997\pm0.0024$&$1.3746\pm0.0020
$&$1.3732$&0.10 \%
                   \\[0pt]\hline
4 &$0.0365\pm0.0005$ & $0.7225\pm0.0022$&$1.4208\pm0.0030
$&$1.4186$&0.15 \%
                   \\[0pt]\hline
5 &$0.0358\pm0.0004$ & $0.7290\pm0.0016$&$1.4305\pm0.0040
$&$1.4319$&0.10 \%
                   \\[0pt]\hline
6 &$0.0354\pm0.0003$ & $0.7301\pm0.0016$&$1.4322\pm0.0040
$&$1.4344$&0.15 \%
                   \\[0pt]\hline
$\infty$
  &$0.0350\pm0.0003$ & $0.7108\pm0.0010$&$1.3993\pm0.0020
  $&$1.3967$&0.19 \%
                   \\[0pt]\hline
\end{tabular}

\vspace{0.3cm}
\noindent
The case $N=\infty$ corresponds to the  Ising  model with equilibrium
magnetic impurities. In this limit the Ising critical exponents $\al$
and $\nu$ are renormalized according to Fisher \cite{11}. The values
of $\eta$, $\nu$, and $\gamma$ were computed by processing original
series for $\eta$, $\nu^{-1}$, and $\gamma^{-1}$, while $\gm_{sc}$
was found by the scaling: $\gm_{sc}=\nu (2-\eta)$. It is seen from
the table that $\gm_{sc}$ differs from $\gm$ by no more than 0.2 \%,
and this may testify the good accuracy of the estimates obtained.

Let us compare our results with earlier estimates of the critical
exponents for the pure Ising model. It is known that due to the
symmetry of the initial Hamiltonian of the cubic model at $N=2$ the
critical
exponents for the cubic and the Ising fixed points coincide. There
were found the following estimates for the Ising model \cite{7}:
$\eta =0.035\pm0.002$, $\nu =0.628\pm0.001$. These numbers are in
excellent agreement with the data of Table 2 at $N=2$. Our estimates
are also in a good accordance  with the results of Ref. \cite{9},
where a substantially different method was employed.

In conclusion let us formulate the results of the present paper.
A new approach to summation of divergent series has been suggested.
The method employes the Borel transformation
combined with a conformal mapping. It relies upon the stability of the
result of processing on the transformation parameters and therefore does not
require to know the exact asymptotic behavior of the series.
The method has been tested on the functions expanded in their asymptotic
series and applied to estimating critical exponent values for the cubic
model. The principal observation is that within the new approach to
summation of the perturbative series of both simple and complex (anisotropic)
models exhibit the same behavior. This allows one to apply the developed
technique to process divergent series arising in a number of anisotropic
models describing phase transitions in real substances \cite{12}. It can be
expected that the new summation method may be useful in other
fields of physics (e.g. QCD and QED) where one deals with divergent series
and conventional resummation techniques are inapplicable.

\newpage
\underline{Figure 1.}
a) Graphs of dependence of  ${\cal F}^i_L(0.5;a,b)$,
$L=13$, on parameter $b$ for some values of parameter $a$.\\
b) Graphs of dependence of ${\cal F}^i_L(g;1,b)$
on parameter $b$ in sixth, ninth, and twelfth orders in $g$
for $g=1$.

\vspace{0.3cm}
\underline{Figure 2.} Convergence of the numerical estimates of the
 function
 $${\cal F}(g)=\int_{-\infty}^{+\infty} e^{-x^2-g x^4}dx$$
 depending on the approximation order $L$ for $g=1$. The
 dashed line corresponds to the exact value  $1.368427...$
 The upper and lower broken lines yield, respectively,
 the upper and lower boundaries for the estimate.
 The best estimate is given by the middle line.

\vspace{0.3cm}
\underline{Figure 3.} Curves demonstrating dependence of the result
 of processing the critical exponent $\gamma^{-1}$
 on parameter $b$ for various values of $a$
 for the cubic model at $N=2$ in five-loop approximation.

\appendix
\section{Appendix}
{Table of the results of processing the asymptotic series
 for the function $${\cal F}(g) = \int_{-\infty}^{+\infty} e^{-x^2-g x^4}dx$$
with the following values of the parameters: $g=0.1$,
$a=a_0=4$, $L=13$, $\la\in [-2, 4]$.
Numbers ${\cal F}^i_{13}(g)$ correspond to different
$\la^i(g;a,b)$ realizing the most rapid convergence of the
processed series.
The exact value is ${\cal F}(0.1)=1.674085856..$
.}

\newpage

\vspace{1cm}
\hspace{-4cm}
\begin{tabular}{|c|c|c|c|c|c|c|c|c|c|}\hline

b  & ${\cal F}^1_{13}(g)$&${\cal F}^2_{13}(g)$&${\cal F}^3_{13}(g)$&
${\cal F}^4_{13}(g)$&${\cal F}^5_{13}(g)$&${\cal F}^6_{13}(g)$&
${\cal F}^7_{13}(g)$&${\cal F}^8_{13}(g)$&${\cal F}^9_{13}(g)$\\\hline
0  & 1.6740858&          &          &          &          &
          &          &          &          \\
1  & 1.6740860& 1.6740860& 1.6740860&          &          &
          &          &          &          \\
2  & 1.6740857& 1.6740857& 1.6740857& 1.6740857& 1.6740857&
 1.6740857&          &          &          \\
3  & 1.6740859& 1.6740859& 1.6740859& 1.6740859& 1.6740859&
 1.6740859& 1.6740859& 1.6740859&          \\
4  & 1.6740859& 1.6740859& 1.6740859& 1.6740859& 1.6740859&
 1.6740859& 1.6740859& 1.6740859& 1.6740859\\
5  & 1.6740859& 1.6740859& 1.6740859& 1.6740859& 1.6740859&
 1.6740859& 1.6740859& 1.6740859& 1.6740859\\
6  & 1.6740859& 1.6740859& 1.6740859& 1.6740859& 1.6740859&
 1.6740859& 1.6740859& 1.6740859&          \\
7  & 1.6740859& 1.6740859& 1.6740859& 1.6740859& 1.6740859&
 1.6740859& 1.6740859& 1.6740858&          \\
8  & 1.6740858& 1.6740859& 1.6740859& 1.6740859& 1.6740859&
1.6740859& 1.6740859& 1.6740858&          \\
9  & 1.6740858& 1.6740859& 1.6740859& 1.6740859& 1.6740859&
 1.6740859& 1.6740859& 1.6740856&          \\
10 & 1.6740858& 1.6740859& 1.6740859& 1.6740859& 1.6740859&
 1.6740858& 1.6740859& 1.6740853&          \\
11 & 1.6740858& 1.6740859& 1.6740859& 1.6740859& 1.6740859&
 1.6740858& 1.6740860& 1.6740846&          \\
12 & 1.6740858& 1.6740859& 1.6740859& 1.6740859& 1.6740859&
 1.6740858& 1.6740862&          &          \\
13 & 1.6740858& 1.6740859& 1.6740859& 1.6740859& 1.6740859&
 1.6740858& 1.6740864&          &          \\
14 & 1.6740858& 1.6740859& 1.6740859& 1.6740858& 1.6740859&
 1.6740857& 1.6740868&          &          \\
15 & 1.6740857& 1.6740859& 1.6740859& 1.6740858& 1.6740859&
 1.6740856& 1.6740873&          &          \\
16 & 1.6740857& 1.6740858& 1.6740859& 1.6740858& 1.6740860&
 1.6740854& 1.6740881&          &          \\
17 & 1.6740857& 1.6740858& 1.6740859& 1.6740858& 1.6740860&
 1.6740852& 1.6740892&          &          \\
18 & 1.6740856& 1.6740858& 1.6740859& 1.6740858& 1.6740862&
 1.6740848& 1.6740906&          &          \\
19 & 1.6740855& 1.6740858& 1.6740859& 1.6740857& 1.6740863&
 1.6740844& 1.6740925&          &          \\
20 & 1.6740855& 1.6740858& 1.6740859& 1.6740856& 1.6740865&
 1.6740838& 1.6740947&          &          \\
21 & 1.6740854& 1.6740858& 1.6740860& 1.6740856& 1.6740868&
 1.6740832& 1.6740978&          &          \\
22 & 1.6740853& 1.6740858& 1.6740861& 1.6740855& 1.6740870&
 1.6740823& 1.6741013&          &          \\
23 & 1.6740852& 1.6740858& 1.6740861& 1.6740852& 1.6740874&
 1.6740814& 1.6741054&          &          \\
24 & 1.6740850& 1.6740858& 1.6740862& 1.6740851& 1.6740878&
 1.6740801& 1.6741106&          &          \\
25 & 1.6740849& 1.6740859& 1.6740863& 1.6740848& 1.6740883&
 1.6740789& 1.6741164&          &          \\
26 & 1.6740871& 1.6740854& 1.6740865& 1.6740844& 1.6740889&
 1.6740771& 1.6741226&          &          \\
27 & 1.6740873& 1.6740853& 1.6740866& 1.6740841& 1.6740896&
 1.6740751& 1.6741285&          &          \\
28 & 1.6740875& 1.6740853& 1.6740869& 1.6740837& 1.6740905&
 1.6740728& 1.6741373&          &          \\
29 & 1.6740878& 1.6740853& 1.6740874& 1.6740833& 1.6740914&
 1.6740702& 1.6741432&          &          \\
30 & 1.6740880& 1.6740852& 1.6740875& 1.6740825& 1.6740926&
 1.6740674&          &          &          \\
\hline
\end{tabular}

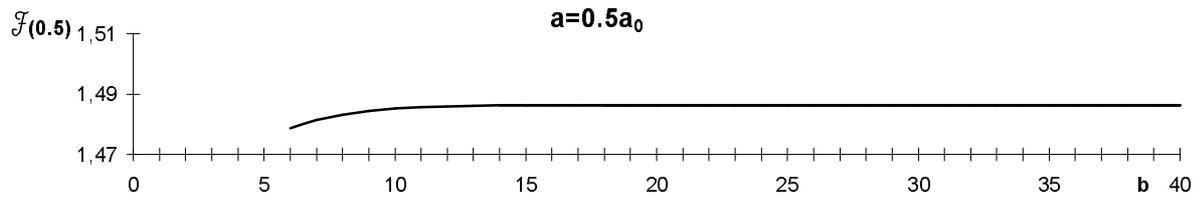
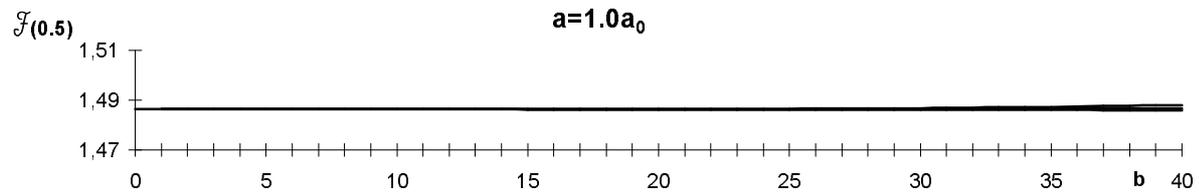
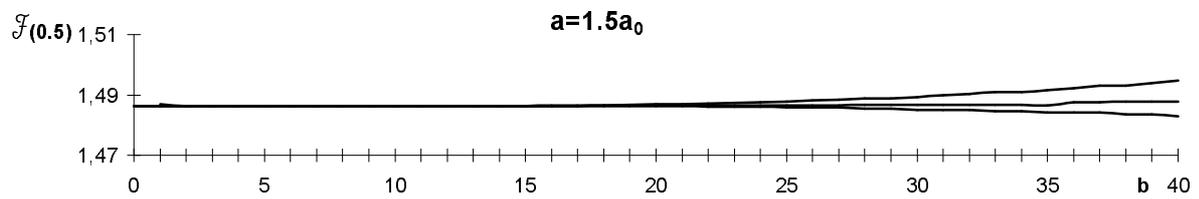
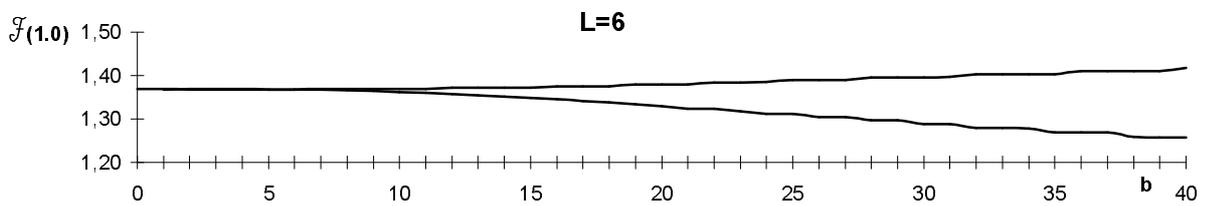
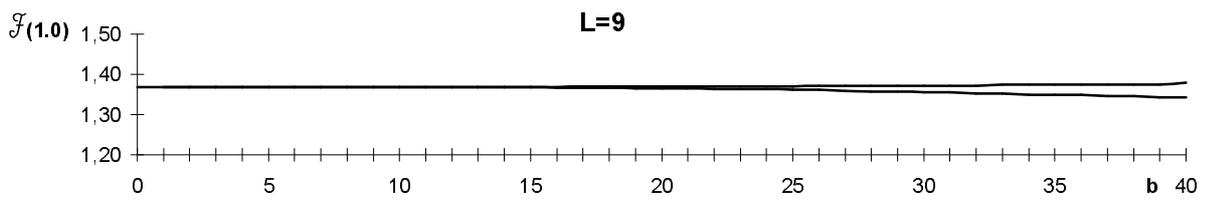
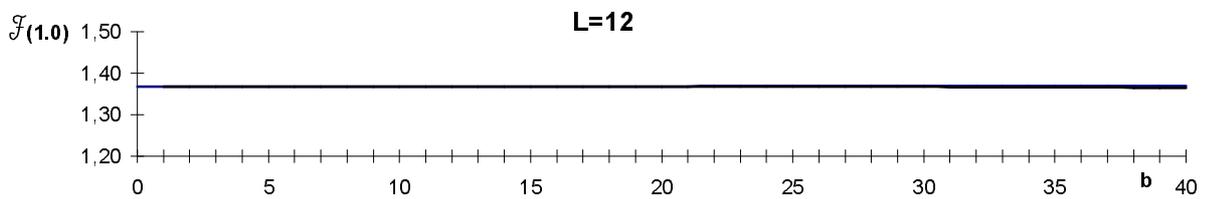
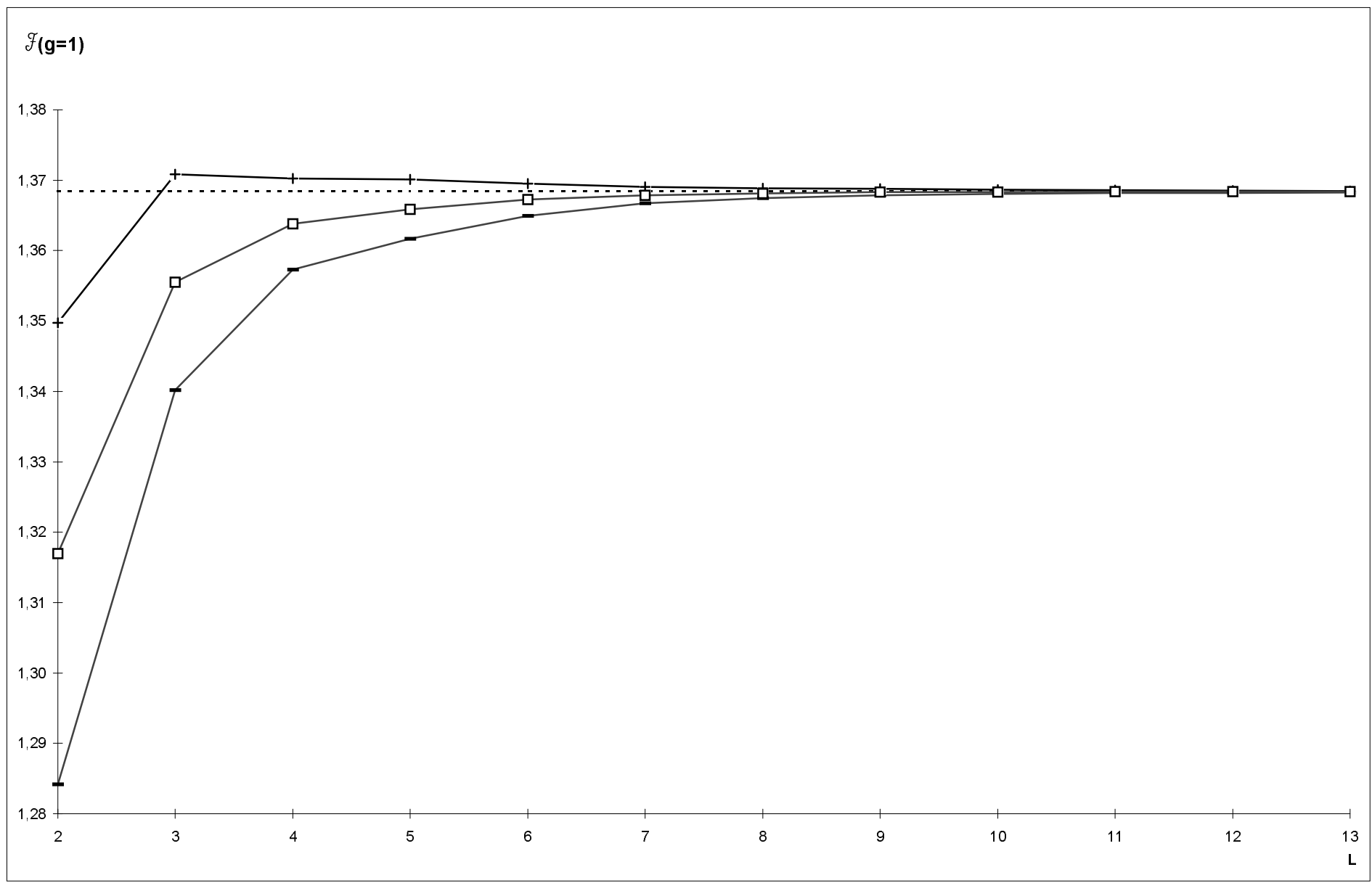
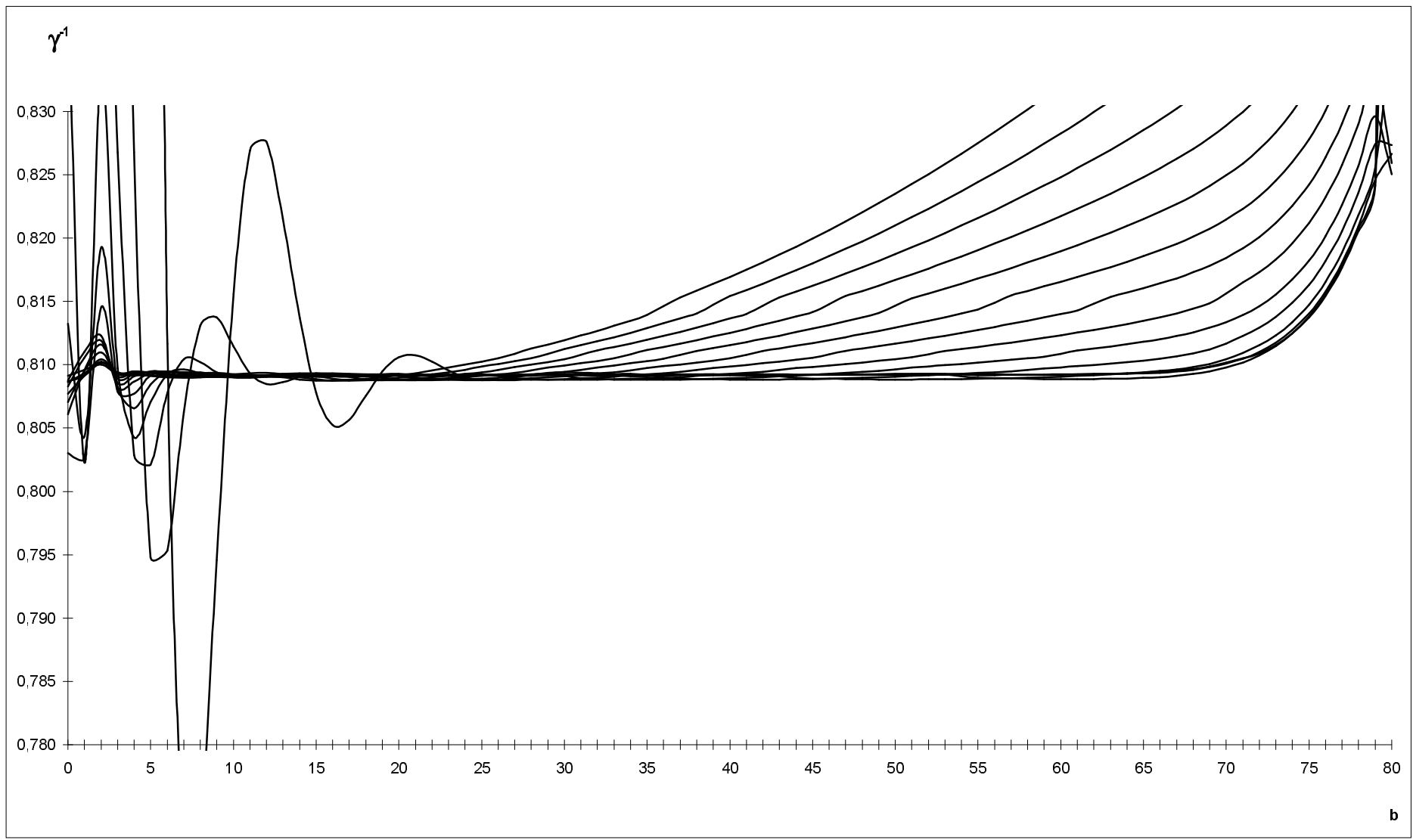
\end{document}